\documentclass[12pt,twocolumn]{article}
\usepackage[latin1]{inputenc}
\usepackage{amsmath,amsfonts,amssymb,graphicx}
\usepackage{authblk}
\usepackage{placeins}
\usepackage{cite}

\begin{document}

\title{
Analysis of structure and dynamics\\ in three-neuron motifs
}

\author[1]{Patrick Krauss}
\author[1]{Alexandra Zankl}
\author[1]{Achim Schilling}
\author[1]{Holger Schulze}
\author[1,2]{Claus Metzner}
\affil[1]{Experimental Otolaryngology, Neuroscience Group, University Hospital Erlangen, Friedrich-Alexander University Erlangen-N\"urnberg (FAU), Germany}
\affil[2]{Department of Physics, Biophysics Group, Friedrich-Alexander University Erlangen-N\"urnberg (FAU), Germany}

\maketitle

Corresponding author: \\
Dr. Patrick Krauss  \\
Experimental Otolaryngology \\
Friedrich-Alexander University of Erlangen-N\"urnberg \\
Waldstrasse 1 \\
91054 Erlangen, Germany \\
Phone:  +49 9131 8543853 \\
E-Mail: patrick.krauss@uk-erlangen.de \\ \\

\textbf{Keywords:} \\
Three-Node Network Motifs, Neural Networks, Boltzmann Neurons, Structure and Dynamics \\ \\ \\

\newpage

\begin{abstract}
In neural networks with identical neurons, the matrix of connection weights completely describes the network structure and thereby determines how it is processing information. However, due to the non-linearity of these systems, it is not clear if similar microscopic connection structures also imply similar functional properties, or if a network is impacted more by macroscopic structural quantities, such as the ratio of excitatory and inhibitory connections (balance), or the ratio of non-zero connections (density). To clarify these questions, we focus on motifs of three binary neurons with discrete ternary connection strengths, an important class of network building blocks that can be analyzed exhaustively. We develop new, permutation-invariant metrics to quantify the structural and functional distance between two given network motifs. We then use multidimensional scaling to identify and visualize clusters of motifs with similar structural and functional properties. Our comprehensive analysis reveals that the function of a neural network is only weakly correlated with its microscopic structure, but depends strongly on the balance of the connections.
\end{abstract}

\newpage

\section*{Introduction}

Recently, a number of projects seek to map the human connectome, aiming to connect its structure to function and behavior\cite{markram2012human, van2013wu, glasser2016human}. However, even if the connectome would be known completely, it remains an unresolved problem how to translate this detailed structural data into meaningful information processing functions and algorithms \cite{jonas2017could}. For instance, the connectome of \textit{Caenorhabditis elegans} is known for decades and is with only 302 neurons orders of magnitudes smaller than the human brain. Nevertheless, even this relatively small system is not yet understood in terms of its dynamics, let alone at a functional level \cite{gray2005circuit, hobert2003behavioral}.

Moreover, the problem is complicated by the fact that very similar dynamics of a neural network at a macroscopic level might be realized by very different structures at the microscopic level \cite{newman2003structure}. Therefore, an important step towards extracting function from structure is a tool to quantitatively compare different structures and dynamics. 

In a neural network, all relevant structural information is encoded in a weight matrix, containing the mutual connection strength of all neurons \cite{hertz1991introduction, schmidhuber2015deep, lecun2015deep, goodfellow2016deep}. Quantifying the similarity of two weight matrices by standard measures, such as the sum of squared differences between corresponding matrix elements, is however not sufficient because of possible permutations of the neuron indices. Similarly, the dynamical properties of a neural network are encoded in a matrix of transition probabilities between all possible network states. As mentioned before, comparing the sum of squared differences between corresponding matrix elements fails in case of neuron permutations. 

To solve this problem, we develop permutation-invariant metrics to quantify, both, the structural and dynamical distance between two given networks. We apply these distance metrics to so-called motifs, a class of small recurrent networks which have been shown to be fundamental building blocks of various complex networks \cite{milo2002network}, such as gene regulatory networks \cite{shen2002network, alon2007network}, the world wide web \cite{milo2002network}, and the human brain \cite{song2005highly}.

We exhaustively compute the structural and dynamical distances between all possible pairs of the $3411$ different classes of three-neuron motifs with ternary connection strengths, resulting in two distance matrices with $3411 \times 3411$ entries each. Based on these matrices, we use multidimensional scaling \cite{kruskal1964multidimensional, kruskal1964nonmetric, cox2000multidimensional, borg2017applied, krauss2018statistical} to visualize the structural and dynamical similarity relations between different motifs on a two-dimensional plane.

Remarkably, it turns out that the distribution of motifs, both in structural and dynamical 'space', is not uniform, but strongly clustered. Moreover, the position of a motif within structural and dynamical space correlates with the ratio of excitatory and inhibitory connections (balance) in the motif's connection matrix. By contrast, no such correlation is found regarding the density (ratio of non-zero connections and all possible connections) of the motif.

\newpage

\section*{Methods}

\subsection*{Three-neuron motifs}
Our study is based on Boltzmann neurons \cite{hinton1983optimal} without bias. The total input $z_i(t)$ of neuron $i$ at time $t$ is calculated as:
\begin{equation}
	\mathrm{z_i(\textit{t})} = \sum\limits_{j=1}^{N} w_{ij}\;y_j(t-1)\,
\end{equation} 
where $y_j(t-1)$ is the binary state of neuron $j$ at time $t-1$ and $w_{ij}$ is the corresponding weight from neuron $j$ to neuron $i$. The probability $p_i(t)$ of neuron $i$ to be in state $y_i(t)=1$ is given by: 
\begin{equation}
	p_i(t) = \sigma(z_i(t)),
\end{equation}
where $\sigma(x)$ is the logistic function
\begin{equation}
	\sigma(x) = \frac{1}{1\;+\;e^{-x}}.
\end{equation}

We investigate the set of all possible network motifs that can be built from 3 Boltzmann neurons with ternary connections $w_{ij}\in\{-1,0,+1\}$, where self connections $w_{ii}$ are permitted (Figure 1a). In principle there are $3^9=19683$ possible ternary $3 \times 3$ weight matrices. However, due to permutation of the neuron indices, not every matrix corresponds to a unique motif class. By exhaustive listing and classification, we find that under these conditions there are exactly 3411 distinct motif classes. For later convenience we label all motif classes with unique indices, which are derived from the corresponding weight matrices. 

\subsection*{State transition matrices of motifs}

Since every neuron can be in one of two binary states, a 3-node motif can be in $2^3=8$ possible motif states. Given the momentary motif state and the weight matrix, the probabilities for all 8 successive motif states can be computed, thus defining the $8 \times 8$ state transition matrix of a Markov process (Figure 1b). All information theoretical properties of 3-neuron motifs, such as entropy or mutual information of succesive states, are determined by the state transition matrix. We therfore calculate the transition matrices for each of the 3411 motif classes.

\subsection*{Motif classes}

A motif class $A$ is defined as the set $\{A^{(m)}:m=1\ldots 6\}$ of weight matrices, which are all related to each other by index permutations, such as $a_{i,j}\rightarrow a_{i,j}^{(m)} = a_{\pi_m(i),\pi_m(j)}$, where $\pi_m$ is the m-th permutation (Figure 2).

\subsection*{Unique labels of motif classes}
The nine entries of the weight matrix $W = \begin{pmatrix}
a&b&c \\
d&e&f\\
g&h&i\\
\end{pmatrix} $ of one motif class are treated as a vector $\mathbf{w} = (abcdefghi)$. the components of this vector are then treated as the digits of a number in the ternary system:
\begin{equation}a\cdot 3^8 + b\cdot 3^7 + c\cdot 3^6 + d\cdot 3^5 + e\cdot 3^4 + f\cdot 3^3 + g\cdot 3^2 + h\cdot 3^1 + i\cdot 3^0 .\end{equation} \label{fternary}
It can be simplified to
\begin{equation} name = \sum _{i=0} ^8 \mathbf{w}[i] \cdot 3^{8-i} .\end{equation}
Here, $\mathbf{w}[0]$ equals the first entry of the vector $\mathbf{w}$ and the value of the sum is the name of the motif. Due to the possible entries $\mathbf{w}[i] \in \{-1,0,1\}$ the motif names range between ``-9841'' and ``9841'', starting with the motif with just ``-1'' as entries and finishing in the motif with just ``1'' as entries. Of course not every number in this range is assigned a motif class as there are in total only 3411 motif classes. This version of the formula is used because the motif class with just zeros as entries gets the name ``0'' and the names are approximately symmetrical around that motif class. Furthermore, in order to make the system more balanced, each motif class is represented by the weight matrix with the smallest absolute value of $name$ among all of its permutations (Figure \ref{fig2})

\subsection*{Structural distance between motif classes}

The structural distance is calculated as follows (Figure \ref{fig3}): Given are two motif classes $A, B$. For each class we derive all 6 permutated weight matrices $A^{(m)}$ and $B^{(n)}$.
For each of the 36 pairs of weight matrices $A^{(m)}$ and $B^{(n)}$, we compute a generalized Hamming distance $\hat{h}$, defined as the number of different ternary matrix elements:

\begin{equation}
	\hat{h}(A^{(m)},B^{(n)}) = \sum_{i,j}\; (1-\delta_{a_{i,j}^{(m)}, b_{i,j}^{(n)}}),
\end{equation}
where $\delta_{x,y}$ is the Kronecker symbol.
The structural distance $d_{str}$ between motif classes matrices $A, B$ is defined as the smallest of 
the above 36 Hamming distannces
\begin{equation}
	d_{str}(A,B)= \mbox{min}_{m,n} \left(\hat{h}(A^{(m)},B^{(n)})  \right)
\end{equation}

\subsection*{Dynamical distance between motif classes}

The structural distance is calculated as follows (Figure \ref{fig4}): For each motif classes $A$, we compute general features $F(A)$, which can be scalars, vectors or matrices. In the case of matrix-like features $F$ and $G$, the Euclidean distance is defined as
\begin{equation}
	d(F,G) =  \sqrt{ \sum_{i,j} (f_{i,j}-g_{i,j})^2 }
\end{equation}
To compute the dynamical distance $d_{dyn}(A,B)$ between two motif classes $A$ and $B$, we derive all 36 pairs of features (e.g. the state transition matrix) from permuted weight matrices $(F(A^{(m)}),F(B^{(n)}))$ and calculate the Euclidean distance $d \left( F(A^{(m)},F(B^{(n)})  \right) $ of each pair. 
The dynamical distance $d_F$ between motif classes $A, B$ is defined as the smallest of the 36 Euclidean distances
\begin{equation}
	d_{dyn}(A,B)= \mbox{min}_{m,n}    \left(\; d (  \; F(A^{(m)},F(B^{(n)})  \; )  \right)
\end{equation}

\subsection*{Multidimensional scaling}
Multidimensional scaling \cite{kruskal1964multidimensional, kruskal1964nonmetric, cox2000multidimensional, borg2017applied, krauss2018statistical} is a statistical method that maps proximity data on pairs of objects (i.e., data expressing the similarity or the dissimilarity of pairs of objects) into distances between points in a multidimensional space so that the pairwise distances follow the given $N \times N$ proximity matrix. We apply multidimensional scaling to visualize the relative positions of motifs in structural and dynamical space, according to their mutual distances $d_{str}$ and $d_{dyn}$ on a two dimensional plane.

\newpage

\section*{Results}

We have shown that there exist $3411$ structurally distinct three-neuron motif classes with ternary connection strengths. Furthermore, we have computed the structural and dynamical distances between all possible pairs of these motif classes, resulting in two $3411 \times 3411$ distance matrices. A scatter plot of all pairwise dynamical and structural distances reveals that there is only a very weak positive correlation between dynamics and structure on a microscopic level (Figure \ref{fig5}).

In a next step, we have used multidimensional scaling \cite{kruskal1964multidimensional, kruskal1964nonmetric, cox2000multidimensional, borg2017applied, krauss2018statistical} to identify and visualize the structural and dynamical similarity relations between motif classes on a two-dimensional plane. Strikingly, we find that the distribution of motif classes in both, structural and dynamical 'space' is not uniform, but instead reveals strong clustering (Figure \ref{fig6}). Furthermore, the structural distribution (Figure \ref{fig6}a, c) reveals a 6-fold rotation symmetry, which might be due to the 6 possible permutations of 3-neuron motifs. 

Although there is no strong correlation between dynamics and structure on a microscopic level, the position of a motif within structural and dynamical space strongly correlates with a macroscopic parameter, namely the ratio of excitatory and inhibitory connections (balance) in the motif's connection matrix (Figure \ref{fig6}a, b). By contrast, no such correlation is found regarding another macroscopic parameter, the density (ratio of non-zero connections and all possible connections) of the motif (Figure \ref{fig6}c, d). 

\newpage

\section*{Discussion}

The relation of structure and function is a long-standing topic in biology \cite{bullock1965structure, estes1989rotavirus, blackburn1991structure, harris1996structure, missale1998dopamine, mitchell2011arterial}. On the one hand, the micro-structure of a biological system determines the set of possible functions that this system can serve. On the other hand, human observers may not be able to deduce the function of a system from its structure alone: even if we know all neural connection strengths in some sub-network of the animal brain, as well as all its input and output signals, the specific purpose of this sub-network within the whole of the organism may remain elusive \cite{jonas2017could, gray2005circuit, hobert2003behavioral}. Indeed, 'function' is not a property of the isolated subsystem alone, but can only be defined in the context of its embedding global system. For this reason, we focus in this work not on the function of neural systems, but on their dynamics - a property that is completely determined by the network structure and, if present, the system's input signals. 

An additional advantage of this approach is that dynamics, just as structure, can be conveniently expressed in the form of matrices. Based on these matrices, we have developed suitable metrics that measure the distance of two neural networks in structural or dynamical space respectively. Using this tool, we can investigate how sensitive network dynamics reacts to small changes in network structure. Robustness with respect to structural changes is crucial in biological brains, as the synaptic weights cannot be adjusted with extremely high accuracy  \cite{pinneo1966noise,  faisal2008noise, rolls2010noisy}.

For the case of isolated three-neuron networks, we have found that the question of robustness has no definitive answer on the microscopic level of individual neuron connection strength: a small change in the connection weights can have, both, small and large dynamical consequences. By contrast, a much clearer correlation is found between certain statistical (macroscopic) properties of a network's weight matrix and its dynamics. In particular, the ratio of excitatory to inhibitory connections affects network dynamics very strongly, while the ratio of non-zero connections is much less important.

This result suggests that biological brains might control the overall statistics of the connection weights in order to drive the neural networks into a desired dynamical regime. Learning would then be just a fine-tuning of the weights with respect to the required information processing task. 

Future work will need to extend our methods to larger neural networks, and to networks with continuous connection strengths between the neurons. In this case, it will no longer be feasible to enumerate all possible neuron permutations in order to compute the minimum structural or dynamical distance. However, we expect that a Monte-Carlo based minimization of the distance will yield sufficiently accurate results.

\newpage

\begin{figure}[h!]
	\centering
	\includegraphics[width=1.0\linewidth]{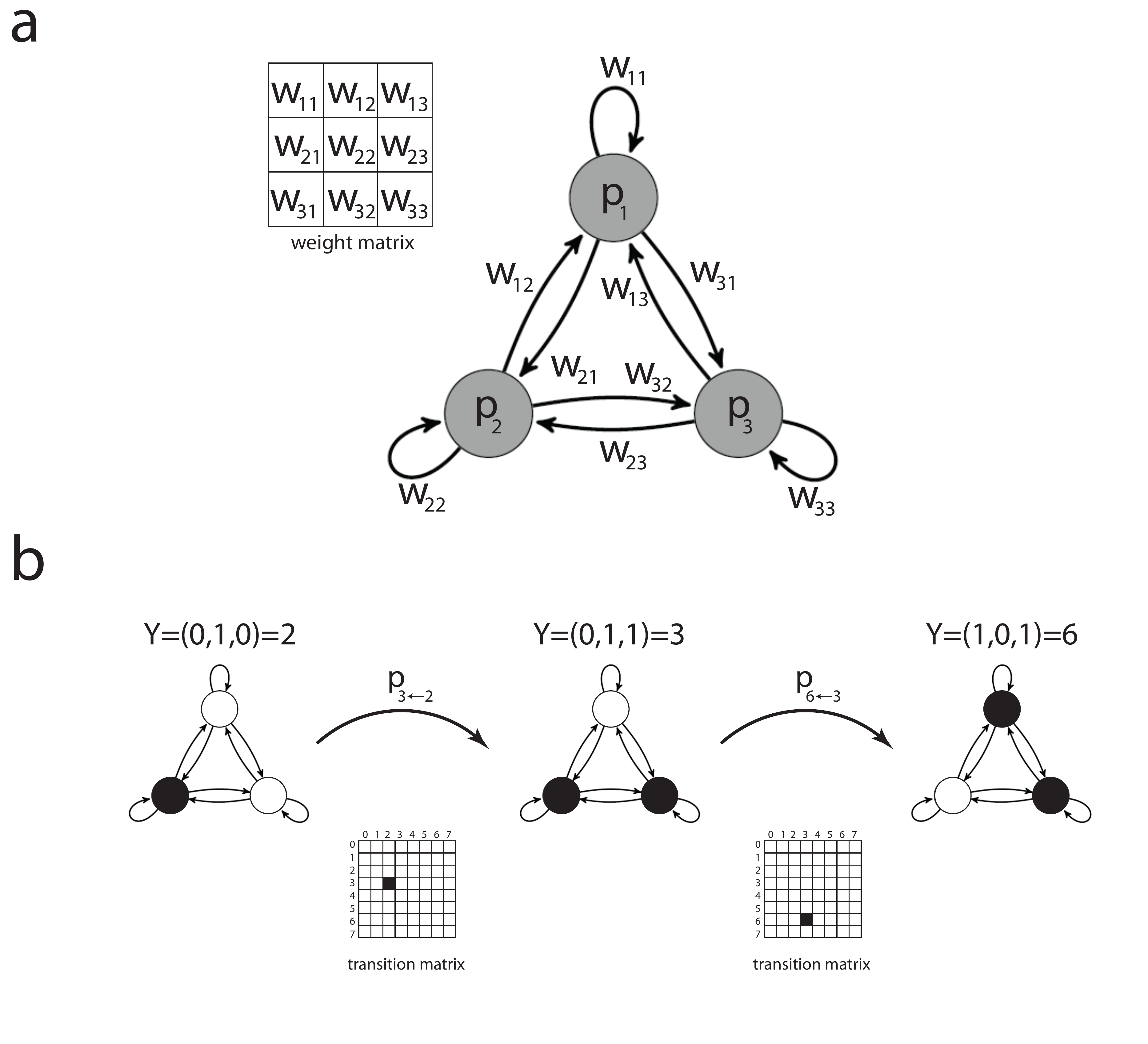}
	\caption{Motifs of three couples Boltzmann neurons. Each motif is characterized by a $3 \times 3$ weight matrix $W$ (a), defining the connection strength between the neurons. There are $2^3=8$ possible states $Y=0  \ldots 7$ for each motif. The transition probabilities between these states are summarized in a $8\times 8$ state transition matrix (b).}
	\label{fig1}
\end{figure}

\begin{figure}[h!]
	\centering
	\includegraphics[width=1.0\linewidth]{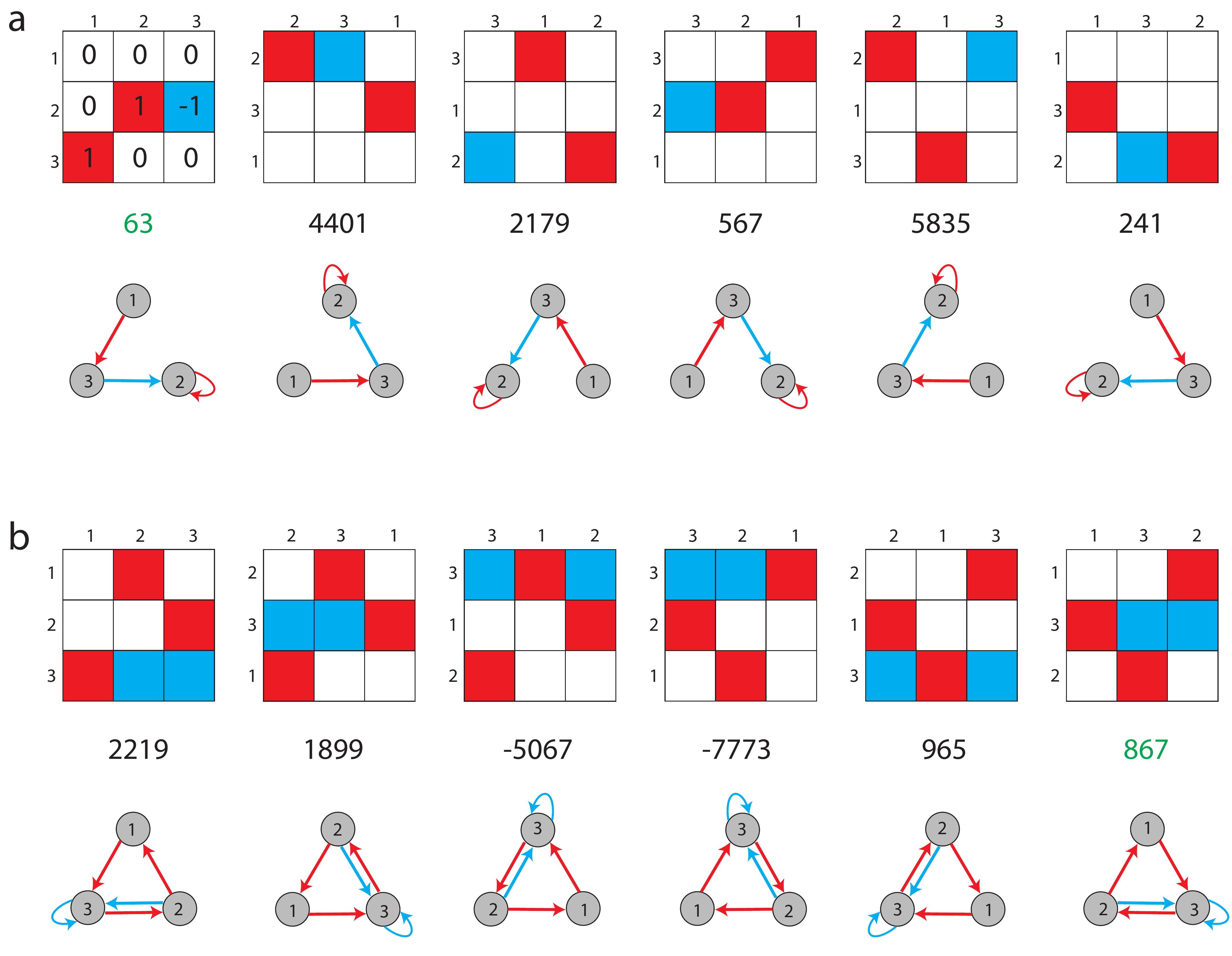}
	\caption{Unique labeling of motif classes. Possible entries in the $3\times 3$ weight matrix of a motif are $-1$ (blue), $0$ (white) and $+1$ (red). Shown are all possible permutations of topologically equivalent motifs for two arbitrary chosen cases (a, b). Each motif class is assigned a unique label (green numbers), as described in the method section.}
	\label{fig2}
\end{figure}

\begin{figure}[h!]
	\centering
	\includegraphics[width=1.0\linewidth]{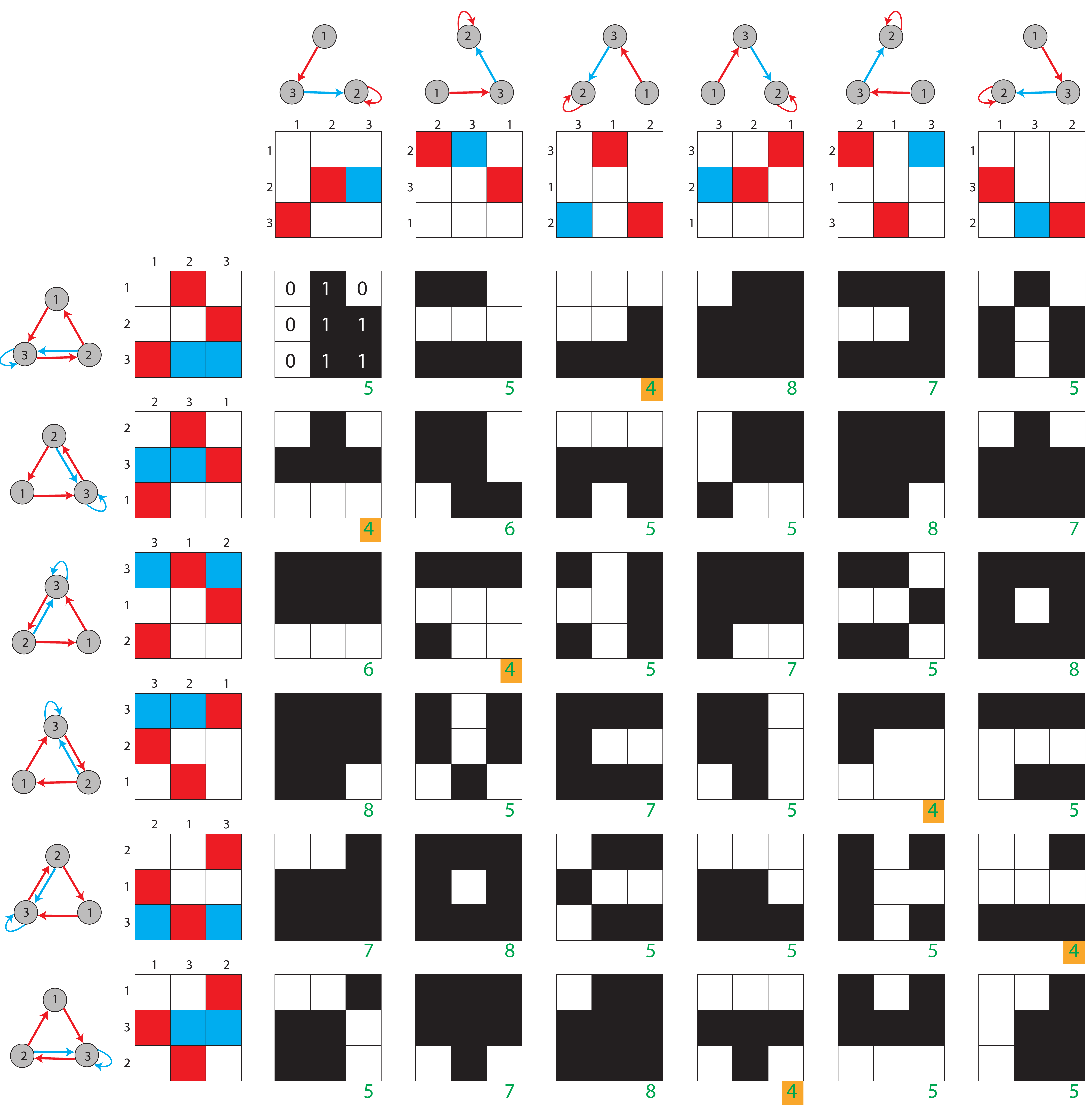}
	\caption{Structural distance between motif classes. Rows and columns show the 6 possible permutations of two given motif classes. For each of the 36 combinations, the generalized Hamming distance $\hat{h}$ (green numbers) is computed. Black and white matrices indicate the Hamming distances between corresponding matrix elements. As described in the method section, the structural distance is defined as the minimum of all 36 generalized Hamming distances (green numbers with yellow background).}
	\label{fig3}
\end{figure}

\begin{figure}[h!]
	\centering
	\includegraphics[width=1.0\linewidth]{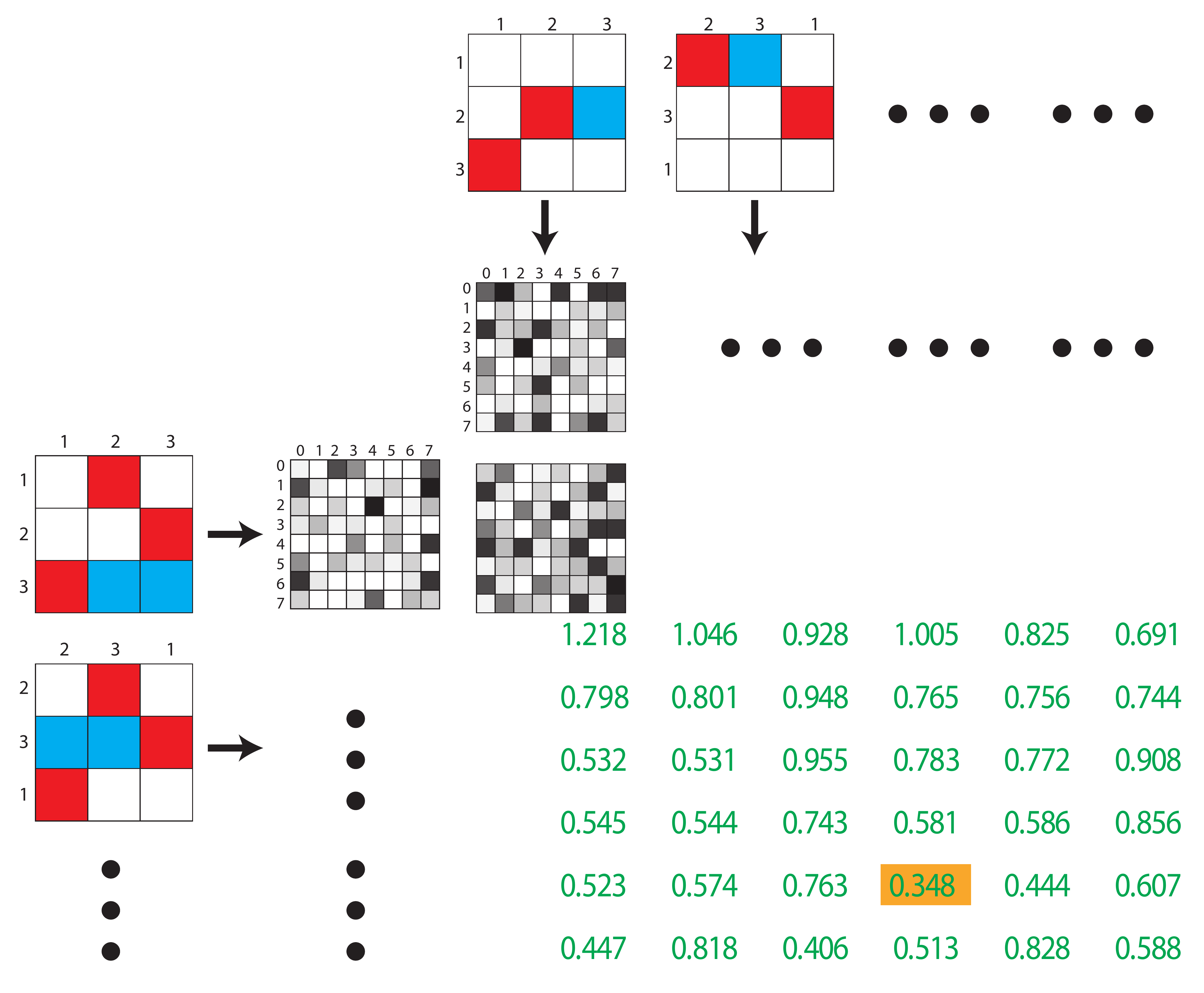}
	\caption{Dynamical distance between motif classes. As in Figure \ref{fig5}, rows and columns contain the 6 possible permutations of two given motif classes. For each permutation the corresponding state transition matrix is calculated (grey shaded matrices). Subsequently, for each of the 36 combinations, the Euclidean distance between each pair of state transition matrices is calculated (green numbers). As described in the method section, the dynamical distance is defined as the minimum of all 36 Euclidean distances (green number with yellow background).}
	\label{fig4}
\end{figure}

\begin{figure}[h!]
	\centering
	\includegraphics[width=1.0\linewidth]{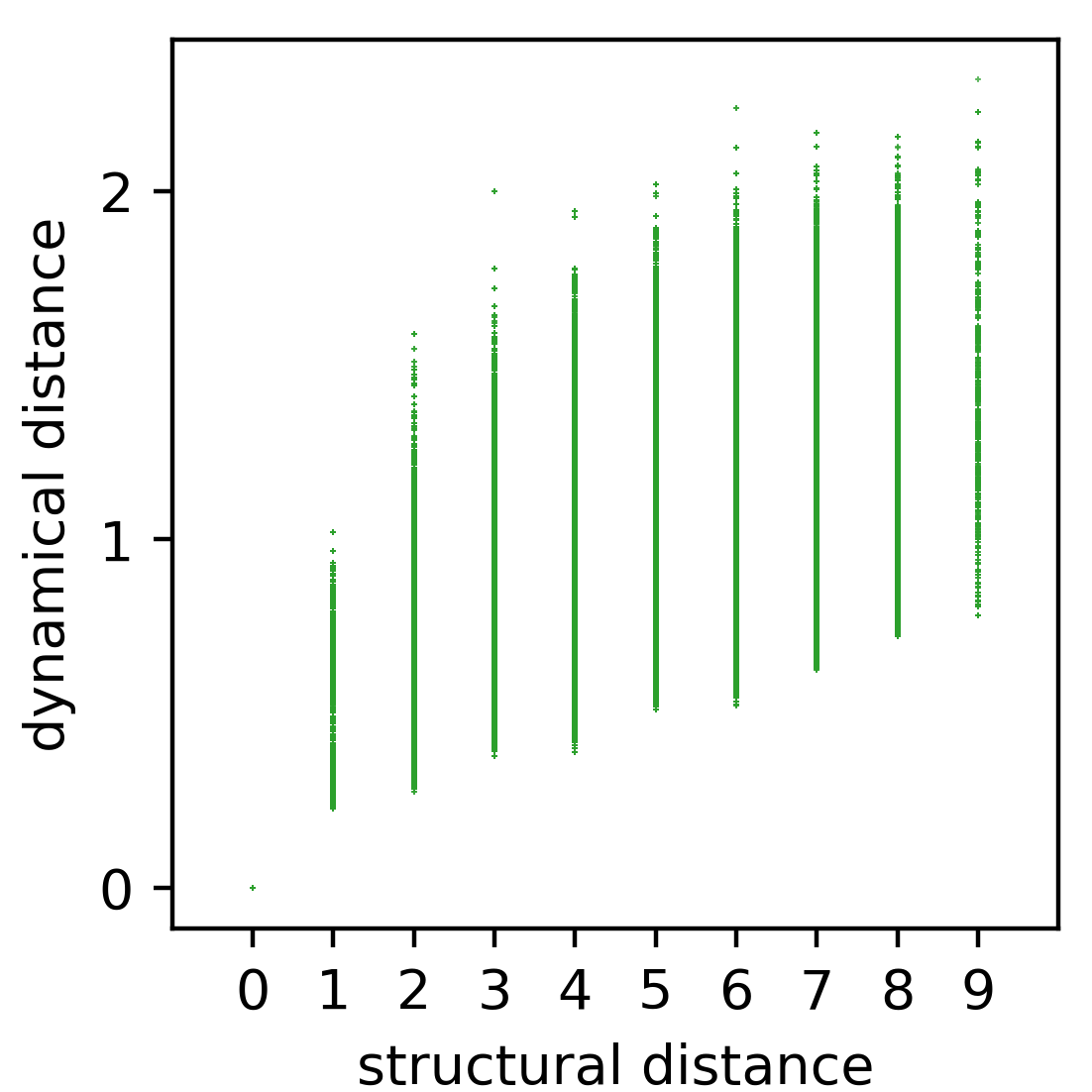}
	\caption{Scatterplot of all pairwise dynamical and structural distances. Each point $(d_{str},d_{dyn})$ represents the relation between structural distance $d_{str}$ and dynamical distance $d_{dyn}$ of the same two motifs. There is only a very weak positive correlation between structure and dynamics.}
	\label{fig5}
\end{figure}

\begin{figure}[h!]
	\centering
	\includegraphics[width=1.0\linewidth]{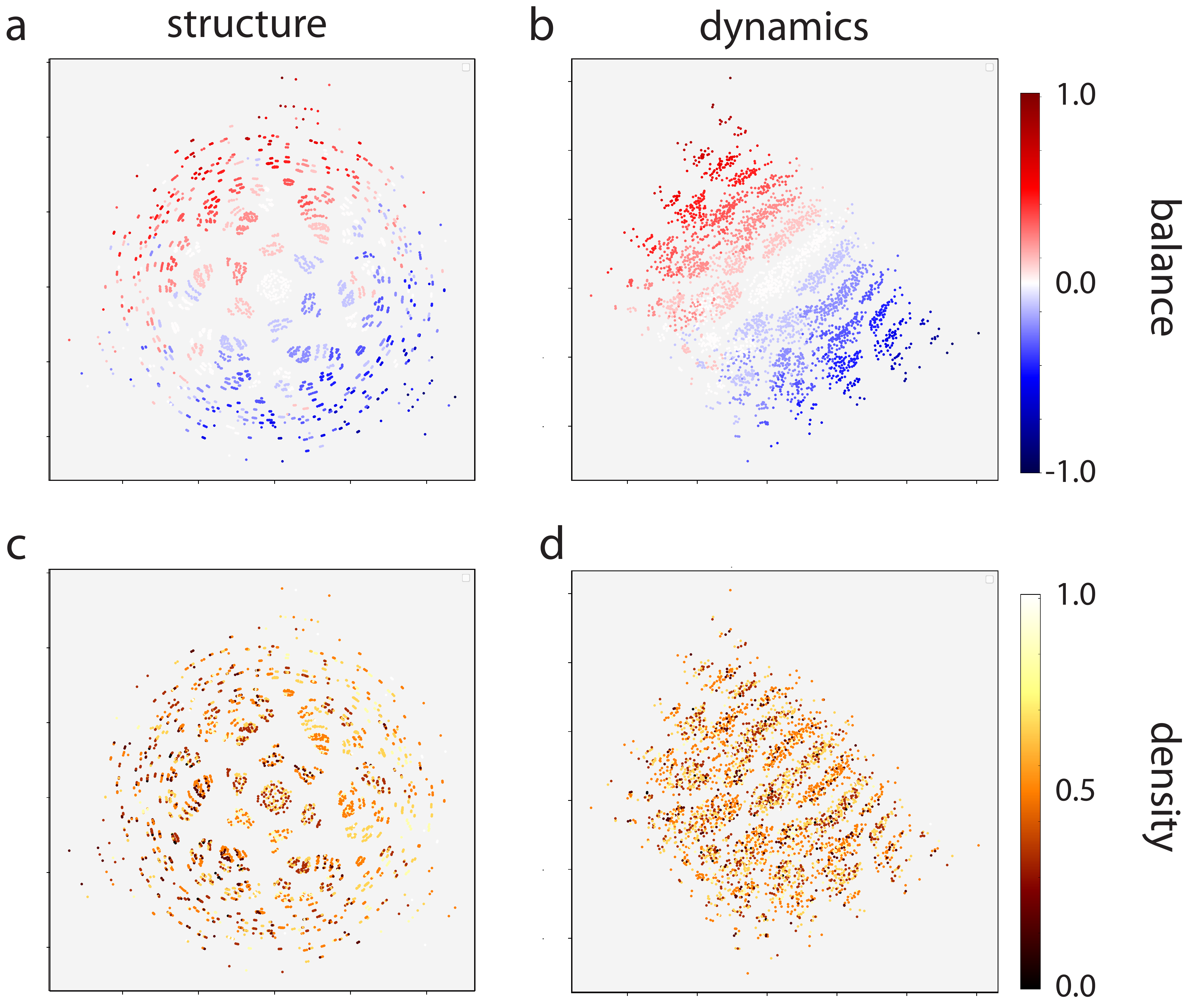}
	\caption{Motif distribution in structural (a, c) and dynamical (b, d) space. Plots are color coded according to balance (a, b) and density (c, d) parameters. The structural distribution reveals a 6-fold rotation symmetry due to the 6 possible permutations of 3-neuron motifs. In addition, motifs are ordered linearly according to the balance parameter, in both structural (a) and dynamical (b) space. By contrast, motifs are not ordered with respect to the density parameter (c, d).}
	\label{fig6}
\end{figure}

\FloatBarrier
\newpage

\section*{Author contributions}
PK and CM designed the study and developed the theoretical approach. AZ performed computer simulations. PK, AS and AZ prepared the figures. PK and CM wrote the paper. HS and AZ provided helpful discussion. All authors read and approved the final manuscript.

\newpage

\section*{Funding}
This work was supported by the Deutsche Forschungsgemeinschaft (DFG, grant SCHU1272/12-1). The authors are grateful for the donation of two Titan Xp GPUs by the NVIDIA GPU Grant Program.

\newpage

\FloatBarrier

\bibliographystyle{unsrt}
\bibliography{refs}

\begin{thebibliography}{10}

\bibitem{markram2012human}
Henry Markram.
\newblock The human brain project.
\newblock {\em Scientific American}, 306(6):50--55, 2012.

\bibitem{van2013wu}
David~C Van~Essen, Stephen~M Smith, Deanna~M Barch, Timothy~EJ Behrens, Essa
  Yacoub, Kamil Ugurbil, Wu-Minn~HCP Consortium, et~al.
\newblock The wu-minn human connectome project: an overview.
\newblock {\em Neuroimage}, 80:62--79, 2013.

\bibitem{glasser2016human}
Matthew~F Glasser, Stephen~M Smith, Daniel~S Marcus, Jesper~LR Andersson,
  Edward~J Auerbach, Timothy~EJ Behrens, Timothy~S Coalson, Michael~P Harms,
  Mark Jenkinson, Steen Moeller, et~al.
\newblock The human connectome project's neuroimaging approach.
\newblock {\em Nature neuroscience}, 19(9):1175, 2016.

\bibitem{jonas2017could}
Eric Jonas and Konrad~Paul Kording.
\newblock Could a neuroscientist understand a microprocessor?
\newblock {\em PLoS computational biology}, 13(1):e1005268, 2017.

\bibitem{gray2005circuit}
Jesse~M Gray, Joseph~J Hill, and Cornelia~I Bargmann.
\newblock A circuit for navigation in caenorhabditis elegans.
\newblock {\em Proceedings of the National Academy of Sciences},
  102(9):3184--3191, 2005.

\bibitem{hobert2003behavioral}
Oliver Hobert.
\newblock Behavioral plasticity in c. elegans: paradigms, circuits, genes.
\newblock {\em Journal of neurobiology}, 54(1):203--223, 2003.

\bibitem{newman2003structure}
Mark~EJ Newman.
\newblock The structure and function of complex networks.
\newblock {\em SIAM review}, 45(2):167--256, 2003.

\bibitem{hertz1991introduction}
John Hertz, Anders Krogh, and Richard~G Palmer.
\newblock {\em Introduction to the theory of neural computation.}
\newblock Addison-Wesley/Addison Wesley Longman, 1991.

\bibitem{schmidhuber2015deep}
J{\"u}rgen Schmidhuber.
\newblock Deep learning in neural networks: An overview.
\newblock {\em Neural networks}, 61:85--117, 2015.

\bibitem{lecun2015deep}
Yann LeCun, Yoshua Bengio, and Geoffrey Hinton.
\newblock Deep learning.
\newblock {\em nature}, 521(7553):436, 2015.

\bibitem{goodfellow2016deep}
Ian Goodfellow, Yoshua Bengio, Aaron Courville, and Yoshua Bengio.
\newblock {\em Deep learning}, volume~1.
\newblock MIT press Cambridge, 2016.

\bibitem{milo2002network}
Ron Milo, Shai Shen-Orr, Shalev Itzkovitz, Nadav Kashtan, Dmitri Chklovskii,
  and Uri Alon.
\newblock Network motifs: simple building blocks of complex networks.
\newblock {\em Science}, 298(5594):824--827, 2002.

\bibitem{shen2002network}
Shai~S Shen-Orr, Ron Milo, Shmoolik Mangan, and Uri Alon.
\newblock Network motifs in the transcriptional regulation network of
  escherichia coli.
\newblock {\em Nature genetics}, 31(1):64, 2002.

\bibitem{alon2007network}
Uri Alon.
\newblock Network motifs: theory and experimental approaches.
\newblock {\em Nature Reviews Genetics}, 8(6):450, 2007.

\bibitem{song2005highly}
Sen Song, Per~Jesper Sj{\"o}str{\"o}m, Markus Reigl, Sacha Nelson, and Dmitri~B
  Chklovskii.
\newblock Highly nonrandom features of synaptic connectivity in local cortical
  circuits.
\newblock {\em PLoS biology}, 3(3):e68, 2005.

\bibitem{kruskal1964multidimensional}
Joseph~B Kruskal.
\newblock Multidimensional scaling by optimizing goodness of fit to a nonmetric
  hypothesis.
\newblock {\em Psychometrika}, 29(1):1--27, 1964.

\bibitem{kruskal1964nonmetric}
Joseph~B Kruskal.
\newblock Nonmetric multidimensional scaling: a numerical method.
\newblock {\em Psychometrika}, 29(2):115--129, 1964.

\bibitem{cox2000multidimensional}
Trevor~F Cox and Michael~AA Cox.
\newblock {\em Multidimensional scaling}.
\newblock Chapman and hall/CRC, 2000.

\bibitem{borg2017applied}
Ingwer Borg, Patrick~JF Groenen, and Patrick Mair.
\newblock {\em Applied multidimensional scaling and unfolding}.
\newblock Springer, 2017.

\bibitem{krauss2018statistical}
Patrick Krauss, Claus Metzner, Achim Schilling, Konstantin Tziridis, Maximilian
  Traxdorf, Andreas Wollbrink, Stefan Rampp, Christo Pantev, and Holger
  Schulze.
\newblock A statistical method for analyzing and comparing spatiotemporal
  cortical activation patterns.
\newblock {\em Scientific reports}, 8(1):5433, 2018.

\bibitem{hinton1983optimal}
Geoffrey~E Hinton and Terrence~J Sejnowski.
\newblock Optimal perceptual inference.
\newblock In {\em Proceedings of the IEEE conference on Computer Vision and
  Pattern Recognition}, pages 448--453. Citeseer, 1983.

\bibitem{bullock1965structure}
Theodore Bullock and G~Adrian Horridge.
\newblock Structure and function in the nervous systems of invertebrates.
\newblock 1965.

\bibitem{estes1989rotavirus}
Mary~K Estes and JEAN Cohen.
\newblock Rotavirus gene structure and function.
\newblock {\em Microbiological reviews}, 53(4):410--449, 1989.

\bibitem{blackburn1991structure}
Elizabeth~H Blackburn.
\newblock Structure and function of telomeres.
\newblock {\em Nature}, 350(6319):569, 1991.

\bibitem{harris1996structure}
Curtis~C Harris.
\newblock Structure and function of the p53 tumor suppressor gene: clues for
  rational cancer therapeutic strategies.
\newblock {\em JNCI: Journal of the National Cancer Institute},
  88(20):1442--1455, 1996.

\bibitem{missale1998dopamine}
Cristina Missale, S~Russel Nash, Susan~W Robinson, Mohamed Jaber, and Marc~G
  Caron.
\newblock Dopamine receptors: from structure to function.
\newblock {\em Physiological reviews}, 78(1):189--225, 1998.

\bibitem{mitchell2011arterial}
Gary~F Mitchell, Mark~A van Buchem, Sigurdur Sigurdsson, John~D Gotal, Maria~K
  Jonsdottir, Olafur Kjartansson, Melissa Garcia, Thor Aspelund, Tamara~B
  Harris, Vilmundur Gudnason, et~al.
\newblock Arterial stiffness, pressure and flow pulsatility and brain structure
  and function: the age, gene/environment susceptibility--reykjavik study.
\newblock {\em Brain}, 134(11):3398--3407, 2011.

\bibitem{pinneo1966noise}
Lawrence~R Pinneo.
\newblock On noise in the nervous system.
\newblock {\em Psychological review}, 73(3):242, 1966.

\bibitem{faisal2008noise}
A~Aldo Faisal, Luc~PJ Selen, and Daniel~M Wolpert.
\newblock Noise in the nervous system.
\newblock {\em Nature reviews neuroscience}, 9(4):292, 2008.

\bibitem{rolls2010noisy}
ET~Rolls and G~Deco.
\newblock The noisy brain.
\newblock {\em Stochastic dynamics as a principle of brain function.(Oxford
  Univ. Press, UK, 2010)}, 2010.

\end{thebibliography}

\newpage

\section*{Conflict of Interest Statement}

The authors declare that the research was conducted in the absence of any commercial or financial relationships that could be construed as a potential conflict of interest.

\end{document}